\documentclass[conference]{IEEEtran}
\usepackage{cite}
\usepackage{graphicx,epsfig,chemarr}
\usepackage{amsmath}

\def\be{\begin{equation}}
\def\ee{\end{equation}}
\def\ba{\begin{eqnarray}}
\def\ea{\end{eqnarray}}
\def\HALF{{\textstyle\frac{1}{2}}}

\def\ie{{\it{i.e.}}}
\def\eai{e_{\alpha i}}
\def\cao{c_{\alpha\omega}}
\def\epsai{\varepsilon_{\alpha i}}
\def\djo{d_{j\omega}}
\def\fa{f_\alpha}
\def\cM{\cal{M}}
\def\cS{\cal{S}}
\def\es{e^\ast}
\def\lamRs{\lambda _{R}^{\ast}}
\def\lamCs{\lambda _{C}^{\ast}}
\def\lamDels{\lambda_{\Delta}^{\ast}}

\begin{document}
\title{The physical language of molecular codes:\\
{A rate-distortion approach to the evolution and emergence of biological codes\\}\bigskip}
\author{\IEEEauthorblockN{Tsvi Tlusty}
\IEEEauthorblockA{Department of Physics of Complex Systems \\
Weizmann Institute of Science \\
Rehovot, Israel 76100\\
Email: tsvi.tlusty@weizmann.ac.il}}
\IEEEspecialpapernotice{ (Invited Paper) \bigskip}
\IEEEoverridecommandlockouts
\IEEEaftertitletext{\center{Submitted to the Workshop on Biological and Bio-Inspired Information Theory,\\ $43^{\text{rd}}$ Annual Conference on Information Sciences and Systems, March 18-20, 2009 \bigskip}}
\maketitle

\begin{abstract}
The function of the organism hinges on the performance of its information-processing networks, which convey information via molecular recognition. Many paths within these networks utilize molecular codebooks, such as the genetic code, to translate information written in one class of molecules into another molecular ``language" . The present paper examines the emergence and evolution of molecular codes in terms of rate-distortion theory and reviews recent results of this approach.

We discuss how the biological problem of maximizing the fitness of an organism by optimizing its molecular coding machinery is equivalent to the communication engineering problem of designing an optimal information channel. The fitness of a molecular code takes into account the interplay between the quality of the channel and the cost of resources which the organism needs to invest in its construction and maintenance. We analyze the dynamics of a population of organisms that compete according to the fitness of their codes. The model suggests a generic mechanism for the emergence of molecular codes as a phase transition in an information channel. This mechanism is put into biological context and demonstrated in a simple example.
\end{abstract}

{\keywords Molecular codes, rate-distortion theory, biological information networks, molecular recognition.}

\section{Introduction}
Molecules are the carriers of information in the living cell. Myriad fluxes of molecular information are produced by the cell's biochemical networks. Other information fluxes enter the cell from the outside environment. All this information is read, integrated and further processed by the circuitry of the cell, which computes the cell's response to this input. This computation often includes the translation of molecular information written in one class of molecules into another class of molecules. For example, genes written in the language of DNA are translated into the language of amino-acids. The translation requires a \emph{molecular code}, in this example, the genetic code \cite{Crick68}. This paper presents and reviews an information-theoretic approach \cite{Tlusty07JTB,Tlusty07Math,Tlusty08PRL,Tlusty08PhysBio,Tlusty08PNAS} (or equivalently, a statistical mechanics approach) to the biological question: How do molecular codes emerge and evolve?

Constructing reliable coding machinery is a challenge to the organism, since this machinery must rely on molecular recognition interactions which take place in the noisy, crowded milieu of the cell. The typical binding energies are not much larger than the energy scale of thermal fluctuations, $k_{B}T$, rendering molecular recognition inherently prone to noise. Moreover, each molecular recognizer needs to locate its correct target within many lookalikes, which further complicates the task of recognition. On top of that, the construction of a code costs the organism time and resources and the organism has to maneuver between the conflicting needs for low cost and high reliability.

To discuss how the interplay between \emph{quality} and \emph{cost} determines the fitness of a molecular code, we describe the code in terms of an information channel or a mapping, which relates two sets of molecules via recognition interactions. One may think of these two sets as molecular ``symbols" and their possible ``meanings". Optimizing molecular codes is a multi-scale task: On the small scale, the accuracy of each recognition event must be maximized (for further details see the conference paper by Y. Savir and \cite{Savir07,Savir08,Savir09}). On the large scale -- which is in the focus of the present paper -- a fitter molecular code should assign meanings to symbols in a manner that reduces the impact of recognition errors, as measured by the \emph{distortion} function.

The need to improve its error-resilience drives the coding machinery to maximal accuracy. However, accurate recognition also requires highly specific binding. We show that the cost of this chemical specificity is equivalent to the \emph{rate} of the molecular information channel, i.e. the mutual information between the symbols and their meanings. The overall fitness of the code is a combination of the rate and the distortion.

As evolution varies the control parameter that measures the relative significance of the rate and the distortion components of the fitness, the organism may reach a point where it becomes beneficial to invest resources in specificity in order to convey information through the channel. At this point -- which is equivalent to a supercritical phase transition in a statistical mechanics system -- a molecular code emerges.

The rest of this paper is organized as follows. In Section \ref{channel}, we define the molecular information channel and the related cost, quality and fitness functions are defined in Section \ref{fitness}. In Section \ref{dynamics}, we derive the critical point, which describes the emergence of the molecular code. We examine a simple example for this generic scenario and discuss it in several regimes of population dynamics. In Section \ref{topology}, we conclude by discussing the effect of topology of the symbol space on the emergence of the code.

\section{Molecular codes as information channels}\label{channel}
Let us consider a molecular code as a mapping between two abstract chemical spaces that contain the two sets of molecules to be related by the code. One may refer to these sets as molecular \textit{symbols} and their respective \textit{meanings}. Perhaps the best known example is the genetic code \cite{Crick68,Tlusty07JTB}, in which the symbols are the 64 codons and the meanings are the 20 amino-acids and the ``stop" signal . A much larger molecular coding system, with thousands of symbols and meanings, is the transcription regulatory network. In this case, the DNA binding sites are the symbols and their potential meanings are the transcription factors that bind the sites \cite{Shinar06,Itzkovitz06}.

\begin{figure}[!t]
\begin{center}
\includegraphics[scale =1, trim = 10mm 120mm 20mm 70mm,clip,width=100mm]{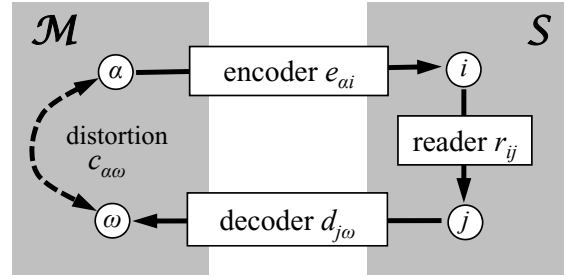}

\caption{\textbf{Molecular codes as noisy information channels. }
A molecular code is a mapping that relates the space $\cM$ of molecular ``meanings", $\alpha, \beta... \omega$ (left), with the space $\cS$ of molecular ``symbols", $i, j, k...$ (right). The noisy communication channel is a three-stage Markov process (solid arrows), where each stage is described its own stochastic matrix (see text): (\emph{i}) A meaning, say $\alpha$, is encoded as a symbol $i$ by the encoder matrix $\eai$.
(\emph{ii}) The symbol $i$ is read as $j$ by the reader matrix $r_{ij}$. (\emph{iii}) The symbol $j$ is decoded as $\omega$ by the decoder matrix $\djo$. The ``distance" between the original meaning $\alpha$ and the reconstructed meaning $\omega$ is given by the matrix element $\cao$ (dashed arrow). The distortion $D$ (\ref{D}) is the average distance $\left< \cao\right>$ along all possible paths $\alpha \rightarrow i \rightarrow j \rightarrow \omega$. The cost $I$ (\ref{I}) is the mutual information between $\cM$ and $\cS$. The linear combination of $D$ and $I$ is the fitness $H$ (\ref{H}).}
\label{f:channel}
\end{center}
\end{figure}

It is evident from the terminology of symbols and meanings  that the problem of optimizing the quality and cost of a molecular code is actually a \emph{semantic }problem: One has to assign meanings to symbols in an optimal manner that maximizes quality while minimizing the cost. To discuss this semantic problem, we consider a two-way information channel that relates the symbols space $\cS$ with its $n_s$ symbols, $i, j, k...$ and the meanings space $\cM$ with its $n_m$ meanings, $\alpha, \beta, \gamma...$ (Fig. \ref{f:channel}). The channel is `two-way' since it describes how meanings are encoded and stored in memory as molecular symbols (the $\cM \rightarrow \cS$ direction), and how the symbols are read and then decoded to reconstruct the meaning (the $\cS \rightarrow \cM$ direction). For a simplified `one-way' formulation see \cite{Tlusty08PhysBio}.

The information channel relies on error-prone molecular recognition and is therefore modeled as a three-stage Markov chain of stochastic processes \cite{Berger71,Luttrell94,Graepel97,Rose98,Tishby99}: (\emph{i}) The representation of meanings as symbols is described by the stochastic encoder matrix $e: \cM \rightarrow \cS$. The matrix element $\eai$ is the probability that a meaning $\alpha$ is encoded by a symbol $i$ (each row obeys probability conservation $\sum_i \eai =1$). (\emph{ii}) Next, the symbol is read. This process is described by the reader matrix $r: \cS \rightarrow \cS$. The matrix element $r_{ij}$ is the probability to read the symbol $i$ as $j$, which accounts for possible misreading errors. The diagonal elements are the probabilities to correctly read the symbols ($\sum_i r_{ij} =1$). (\emph{iii}) Finally, the read symbol is decoded according to the decoder matrix $d: \cS \rightarrow \cM$. The matrix element $d_{j\omega }$ is the probability that the symbol $j$ is interpreted as carrying a meaning $\omega $ ($\sum_\omega \djo =1$). 
The original meaning $\alpha $ passes the three stages of the channel and returns as a reconstructed meaning $\omega $. The effect of errors in the channel is measured by the distance $c_{a\omega }$. Returning to the example of the genetic code, the channel encodes amino-acids as DNA codons. These codons are in turn read by the anti-codons of the tRNA. At its other side, the tRNA is charged with an amino-acid which is the decoded meaning at the output of the channel. This decoded amino-acid is ligated to the protein which is being synthesized by the ribosome \cite{Crick68,Tlusty07JTB}.

The channel is defined by the pair of stochastic maps, $e$ and $d$. When the cost of constructing a coding machinery is too high, the relation between symbols and meanings is non-specific. At this \emph{non-coding} state, any molecular meaning is equally likely to be encoded by any of the available symbols and the encoder matrix therefore does not depend on the meaning, $\eai= u_i$. Similarly, the decoder matrix at the non-coding state does not depend on the symbol $\djo= f_\omega$, where $f_\omega$ is the \emph{demand} for the meaning $\omega$, which accounts for the possibility that certain meanings are used more frequently than others. As will be discussed below, a code emerges when the matrices $e$ and $d$ become non-uniform. The non-uniformity signifies preference for binding between certain molecular symbols and meanings.

\section{The fitness of molecular codes}\label{fitness}

After we defined molecular codes in terms of noisy information channels, we derive the quality and cost of these channels and use them to construct the fitness of the code.
\subsection{The distortion measures the quality of the code}
A natural way to estimate the \emph{quality} of a molecular code is by examining how well the meaning that is reconstructed at its output preserves the  original meanings at the input of the channel. This is measured by the \emph{distortion} function $D$ \cite{Shannon59,Berger71,Cover06}, which is the average distance $\langle \cao \rangle$ along all possible paths between original and reconstructed meanings (\cite{Tlusty07JTB,Tlusty08PRL} and references therein). Each of the possible paths $\alpha \rightarrow i \rightarrow j \rightarrow \omega$ is weighted by its probability, $P_{\alpha i j \omega}$, which is the product of the relevant entries in the encoder, reader and decoder matrices and the demand $\fa$ for the original meaning, $P_{\alpha i j \omega}=  \fa\eai r_{ij} \djo \cao$. The resulting distortion function is the trace,
\be
D  = \langle \cao \rangle
     = \sum\limits_{\alpha,i,j,\omega} P_{\alpha i j \omega} \cao
     = \sum\limits_{\alpha ,i,j,\omega}\fa \eai r_{ij}\djo\cao ~.
\label{D}
\ee

The reader matrix $r$ determines the topology of the symbols space $\cS$. It implies some notion of proximity which may be represented as a graph $G(r,\cS)$ whose nodes are the symbols and edges connect symbols that are likely to be confused, \ie , have a significant $r_{ij}$ value \cite{Tlusty07JTB,Tlusty07Math}. Similarly, the distance matrix $\cao$ represents the topology of the meaning space $\cM$. In the case of the genetic code, for example, the reader tends to confuse similar codons that differ by one base only. In the meanings space, close-by amino-acids are those that have similar chemical characteristics, such as polarity or size.
Both topologies affect the distortion function (\ref{D}).

An ``ideal" perfect reader, $r_{ij}=\delta _{ij}$,  would enable the ``organism" to decode as many meanings as there are available symbols, since there is no chance to confuse between symbols. However, for a realistic reader, which is imperfect due to the inherent recognition noise, it is preferable to  decode fewer meanings and thereby minimize the effect of misreading. Moreover, the distortion function drives the preferable codes to be  \emph{smooth}, in the sense that symbols that are likely to be confused encode similar meanings. In other words, the mappings $e:\cM\rightarrow\cS$ and $d:\cS\rightarrow\cM$ tend to be continuous
\cite{Tlusty07JTB,Itzkovitz06,Shinar06,Tlusty07Math,Tlusty08PRL,Tlusty08PNAS,Tlusty08PhysBio}.

\subsection{The rate measures the cost of the code}
Molecular codes utilize physicochemical binding interactions to relate symbols and meanings. The encoder and decoder matrices are, in fact, the binding probabilities between molecules from $\cM$ and $\cS$. A coding system with high binding specificity can accurately read symbols and thereby reduce the chance of assigning the wrong meaning due to misreading errors. It is evident, however, that highly specific binding also costs higher binding energy, which in general necessitates larger binding-sites. The cost of replicating, transcribing and translating the gene segment that encodes the binding site and the cost of keeping this segment free of mutations, are all expected to be roughly proportional to the binding site size. A reasonable estimate for the cost of the code is therefore the average size of the binding sites, which is roughly proportional to the average binding energy.

To estimate the cost, we extract the average binding energy from the encoder matrix $e$. The matrix element $\eai$ is the probability that the molecule carrying the meaning $\alpha$ binds the molecular symbol $i$. For example, in the transcription regulatory network, $\alpha $ may be one of the transcription factors and $i$ is a prospective DNA binding site where $\alpha$ may bind. If the binding and unbinding events are fast, they obey the Boltzmann equilibrium distribution, $\eai \sim \exp \epsai $, where the binding energy $\epsai$ is measured in $k_{B}T$ units.  It follows that the binding energy $\epsai$ scales like the logarithm of the binding probability, $\epsai  \sim \ln \eai$. As a result, the average size of the binding site, and therefore the cost $I$ of the molecular code, are proportional to the average logarithm of the encoder matrix,
\ba
I &=& \sum_{\alpha ,i}\fa\eai \ln \frac{\eai}{u_{i}}
\nonumber\\
  &=&  \sum_{\alpha,i} \fa \eai (\epsai-\bar{\varepsilon}_{i})
     =  \langle \epsai-\bar{\varepsilon}_{i}\rangle ~.
\label{I}
\ea
The reference energies, $\bar{\varepsilon}_{i}=\ln \sum_{\beta }f_{\beta }\exp \varepsilon _{\beta i}$, and the normalization of $\eai$ by $u_i$ ensure that the cost vanishes when the binding is non-specific at the non-coding state, $\eai= u_i$.

 The cost (\ref{I}) is nothing else than the \emph{mutual information} between the symbols and the meanings, which is the entropy reduction due to the symbol-meaning correlation in the encoder. This is a common measure for the cost of a coding system, which measures the average number of bits required to encode one meaning, \ie, the \emph{rate} of information passing through the channel. In principle, one would need to consider also the bit rate of the decoder, $d$. However, the optimal encoder and decoder are related through the Bayes' theorem (see \cite{Tlusty08PRL} and references therein),
\be
\djo \sum\limits_{\beta ,i}f_{\beta }e_{\beta i}r_{ij}
=f_{\omega}\sum\limits_{i}e_{\omega i}r_{ij} ~.
\label{Bayes}
\ee
Relation (\ref{Bayes}) expresses
the intuitive notion that if the encoded meaning $\omega $ is likely to be read as the symbol $j$ then the symbol $j$ tends to be decoded as $\omega $. It also implies that, in practice, it is enough to specify only one of the encoder and decoder matrices in order to characterize the coding machinery of an organism.

\subsection{The fitness is a rate-distortion functional}
To optimize the molecular coding apparatus, its two determinants, the cost $I$ and distortion $D$ must be balanced. For the sake of simplicity, we express this interplay as the maximization of an overall \textit{code fitness}, which is their linear combination
\be
H=-D-\kappa ^{-1}I ~.
\label{H}
\ee
The minus signs reflect the fact that while $I$ and $D$ need to be minimized, the overall fitness $H$ is driven by evolution towards maxima. The coefficient $\kappa =-\partial I/\partial D$ is the \emph{gain}, which measures the bits of information required to decrease the distortion.
The gain $\kappa$ is expected to increase with the richness of the environment and the complexity the organism: A more complicated environment transmits more signals which  require heavier computation of the cell's response to these signals. Similarly, the circuitry of a complex organism requires higher fluxes of information transfer. It is therefore beneficial for this organism to pay a larger cost to improve the quality of its code and thereby reduce the distortion $D$, since it gains more from such an improvement.

Since the cost or the rate is (up to a factor) the entropy loss due to coding, the conjugate parameter $\kappa^{-1}$ is equivalent to a temperature. The distortion $D$ is equivalent to an interaction energy. The combination of rate and distortion is the fitness function $H$,  a ``free energy" which the organism tries to maximize by optimizing the channel parameters \cite{Shannon59,Berger71,Tlusty07JTB,Tlusty08PRL,Tlusty08PhysBio,Tlusty07Math,Tlusty08PNAS}. High ``temperatures" or small gains indicate a rising cost of binding sites, which drives the encoder and the decoder to homogeneity by reducing the specificity of the underlying binding  interactions.  At the other extreme, high gains or low ``temperatures" drive the coding matrices to a non-random inhomogeneous state.  The optimal code $\es$ (and $d^\ast$), the one which maximizes the fitness $H$, is a function of three determinants: the reading matrix $r$, the distance $c$, and the gain $\kappa$. Below we discuss how a molecular code evolves as these determinants are varied. In particular, we show that a molecular code emerges at a critical transition in the noisy information channel.

\begin{figure}[!t]
\begin{center}
\includegraphics[scale =1, trim = 0mm 60mm 0mm 70mm,clip,width=100mm]{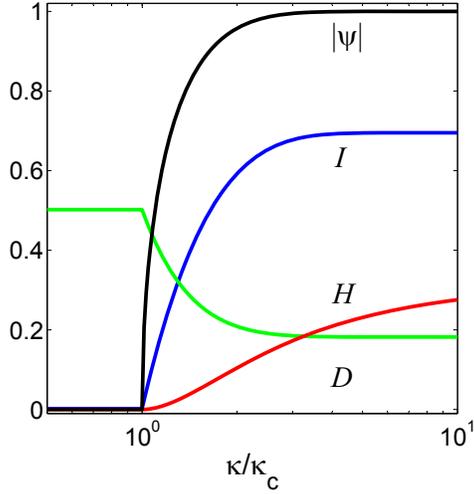}

\caption{\textbf{A simple 2x2 molecular code. }
The code maps two meanings, $\cM = \{\alpha,\omega\}$,  to two symbols, $\cS = \{$\textit{i, j}$\}$. The demand for the two meanings is symmetric $\fa=f_\omega = \HALF$. By probability conservation and symmetry, the encoder has only one degree of freedom and the order parameter $\psi$ is one-dimensional (\ref{simple}). The distance is specified by the parameter $c_0$, the penalty for confusing meanings, whereas the reading matrix $r$ is specified by $\epsilon/2$, the average misreading probability. The evolution of the 2-by-2 code is graphed as a function of the normalized gain $\kappa/\kappa_c$: Plotted are the cost $I$ (blue), the distortion $D$ (green), given by (\ref{IDsimple}). Also plotted are the fitness $H = - D-\kappa^{-1}I$ (red), which is shifted by $c_0/2$, and the order-parameter $\psi$ (black). At low gain (left) the system is in the non-coding state of uncorrelated symbols and meanings ($\psi  = 0$). When $\kappa$ increases above a critical value $\kappa_c = [c_0(1-\epsilon)^2]^{-1}$, the system undergoes a second-order coding transition.  Following the coding transition the cost $I$ increases, but this is compensated by the decreasing distortion $D$, and thus the overall fitness H increases. The parameters are $\epsilon =1/5$, $c_0=1$, which yield $\kappa_c = 25/16$ (\ref{kappac0}).}
\label{f:simple}
\end{center}
\end{figure}

\section{Population dynamics in the code space and the emergence of molecular codes} \label{dynamics}

\subsection{The code  space}
To examine the response of a coding system to changes in the control parameters, $r$, $c$ and $\kappa$, let us consider a population of ``organisms", \ie,  self-replicating information processors that utilize coding systems. The organisms live in an environment where they compete according to the fitness of their codes. The code of each organism is specified by its encoder $e$ and decoder $d$ matrices. However, as discussed above, due to Bayes' theorem it suffices to specify only one of the matrices, say $e$. One may therefore describe the evolution of this population as the motion of points in a \emph{code space} which is spanned by all possible encoders, $0\leq \eai \leq 1$. This space is an $n_m \times n_s$-dimensional unit hypercube.  Each axis of the cube corresponds to one entry of the encoder $\eai$. In fact, since the every row $\alpha$ of the encoder satisfies probability conservation, $\sum_{i}\eai=1$, the effective dimension is reduced to $n_m\times (n_s-1)$.
Each organism is represented by a point in the cube at a location that corresponds to its code. The population is represented by the probability density (or the number density)   $\Psi (\eai)$, which is the probability that a randomly picked organism has a given code.

\subsection{The optimal code}
For the sake of simplicity, we first examine large populations with negligible mutation rate. Such populations peak very sharply around an optimal code $\es$  and can therefore be approximated by a delta distribution  $\Psi \sim \delta (e-\es )$. The dynamics in this regime may be described by the motion of the optimum $\es$ in response to changes in the system parameters, $r$, $c$ and $\kappa$.

The optimal code maximizes the overall fitness $H$ (\ref{H}). To calculate the corresponding encoder $\es$, one augments $H$ with Lagrange multipliers to ensure that the $n_m$ probability conservation relations are satisfied, $H_T = H + \sum_{\alpha}\mu_{\alpha}\sum_{i} \eai$. The optimal encoder code-matrix $\es$ is located at the extremum, $\partial H_T/\partial \eai = 0$ , which leads to \cite{Tlusty08PRL}
\be
\eai^\ast = \frac{ u_i \exp \left(-\kappa \Omega_{\alpha i}\right)} 
                        {\sum\limits_{j} u_j \exp \left(-\kappa \Omega_{\alpha j}\right)} ~.
\label{optimum}
\ee
This is a Boltzmann partition with effective ``energies" $\Omega _{\alpha
i}=\sum\nolimits_{j,\omega }r_{ij}\djo(2 \cao-\sum\nolimits_{\gamma }d_{j\gamma}c_{\gamma\omega})$ and a ``temperature"  $\kappa^{-1}$. Organisms with lower $\kappa $ are ``hotter" and their codes are noisier.

Since both sides of (\ref{optimum}) depend on the code matrix $\es$, through (\ref{Bayes}) and the definition of the $\Omega$-s, it defines a self-consistency relation for $\es$, which in general requires an iterative numerical solution \cite{Tlusty08PRL,Tlusty08PNAS,Cover06}. At low gains, when it is essential for the organism to minimize the cost $I$, the optimal code given  by the solution of (\ref{optimum}) is completely non-specific, $\eai^\ast = u_i$. At this non-coding state $I$ vanishes (\ref{I}) since the encoder conveys no information about the meanings (it is $\alpha$-independent). As we show below, when the gain $\kappa$  increases, the code may remain non-specific for some range of $\kappa$. Then, when it surpasses a certain critical value, $\kappa_c$, the code undergoes a ``coding transition" when it becomes specific.

\subsection{A simple 2x2 code}\label{transition}
To demonstrate the coding transition, we examine the simplest non-trivial example of a coding system that maps two meanings, $\cM = \{\alpha,\omega\}$,  to two symbols, $\cS = \{$\textit{i, j}$\}$ (Fig. \ref{f:simple}). We assume that the demands for $\alpha$ and $\omega$ are equal $\fa = f_\omega = \HALF$, from which it follows that the usage of symbols is also symmetric, $u_i = u_j = \HALF$. The encoder $e$ has four entries and is constrained to a 2D  unit square by the two conservation relations, $\eai +e_{\alpha j}=e_{\omega i}+e_{\omega j}=1$. The symmetry of the setting implies that the distance $c$, the reading $r$, and the encoder $e$,  are all $2$-by-$2$ matrices determined by a single degree of freedom,
$$
r = \begin{pmatrix}
  1-\HALF\epsilon & \HALF\epsilon \\
  \HALF\epsilon & 1-\HALF\epsilon \\
\end{pmatrix}, \quad
c = c_0 \begin{pmatrix}
  0 & 1 \\
  1 & 0 \\
\end{pmatrix},
$$
\be
\text{and}\quad
e = \HALF \begin{pmatrix}
  1+\psi & 1-\psi \\
  1-\psi & 1+\psi \\
\end{pmatrix},
\label{simple}
\ee
The parameter $c_0$ measures the average penalty of replacing $\alpha$ by $\omega$ or $\omega$ by $\alpha$ and $\epsilon/2$ is the average misreading probability. The deviation of the encoder from the uniform non-coding state $\eai=\HALF$ is measured by the order parameter, $-1 \le \psi \le 1$. From (\ref{Bayes}) we find the decoder $d$ and substitution in   (\ref{D}-\ref{I}) yields the distortion and the cost,
\ba
D &=& \HALF c_0 \left[1 - (1-\epsilon)^2\psi^2 \right] \quad \text{and} \nonumber \\
I &=& \HALF \left[ (1+\psi)\ln(1+\psi) + (1-\psi)\ln(1-\psi) \right] ~.
\label{IDsimple}
\ea

Interestingly, the resulting fitness $H = -D-\kappa^{-1}I$ is completely analogous, up to a minus sign, to the free energy of a mean-field Ising magnet. The distortion $D$ corresponds to the spin-spin interaction energy, whereas the cost $I$ corresponds to the entropy of the magnet. Within this analogy, the gain $\kappa$ is the inverse temperature and the magnetic interaction strength $J$ is $J = c_0 (1-\epsilon)^2$. Just like in the magnet, one can increase the order and the correlation in the coding system by raising the interaction strength J -- via increasing $c_0$, or by decreasing the error probability $\epsilon$.
Given the coding system parameters, $c_0$, $\epsilon$, and $\kappa$, one may locate the order-parameter $\psi^\ast$ which maximizes $H$ and determines the optimal code $\es$. The optimum may be found by calculating the extremum, $\partial H/\partial \psi = 0$, or directly from (\ref{optimum}), which yield the familiar self-consistency equation of the Ising magnet
\be
\psi^\ast = \tanh \left[ \kappa c_0 (1-\epsilon)^2 \psi^\ast \right] ~.
\label{psis}
\ee

As in the Ising model, it follows from the solution of (\ref{psis}) that the code remains in the random non-coding state, $\psi^\ast = 0$, as long as the gain $\kappa$ is below a critical value $\kappa_c$, which is equal to the inverse interaction strength,
 \be
 \kappa_c = J^{-1} = [c_0 (1-\epsilon)^2]^{-1} ~.
 \label{kappac0}
 \ee 
 At $\kappa_c$, a coding state, $\psi^\ast \neq 0$, emerges at a continuous, second-order coding transition. Relation (\ref{kappac0}) indicates three possible pathways to approach the coding transition: (i) improving the reading accuracy (smaller $\epsilon$), for example by increasing the size of the  specific binding sites (ii) increasing the penalty $c_0$ of encoding a wrong meaning, and (iii) lowering the importance of cost, by increasing the gain $\kappa$. The first two pathways are equivalent to strengthening the magnetic interaction while the third one is analogous to lowering the ``temperature", $\kappa^{-1}$. A biological example for a possible 2-by-2 molecular code is discussed in \cite{Tlusty08PhysBio}.

\subsection{The critical coding transition}
The notion of a coding transition, demonstrated above in the simple 2-by-2 code, can be generalized to larger coding systems. To locate the coding transition, one examines the stability of the non-coding state, $\eai = u_i$, with respect to small variations of the encoder $\delta \eai = \eai - u_i$. The order parameter $\delta \eai$ reflects the preference of the symbol $i$ to encode the meaning $\alpha$  relative to the average usage $u_i$. This is equivalent to expanding the two-state Ising magnet into an $n_s$-state Potts model. A coding state emerges exactly at the point when the order-parameter becomes non-zero, $\delta\eai \neq 0$, when meanings and symbols become correlated. The coding\slash non-coding transition takes place when the fitness maximum at the non-coding, symmetric state becomes unstable. By analysis of the curvature of the fitness landscape $H(e)$ \cite{Tlusty08PRL}, we find that the critical gain is
\be
\kappa _c = \frac{1}{2\lamRs\lamCs} ~,
\label{kappa_c}
\ee
where $\lamCs$ is the maximal eigenvalue of the normalized distance, $C_{\alpha \omega}=(\fa f_\omega)^{1/2} (\sum_{\beta
}f_\beta c_{\beta \omega}+\sum_{\gamma}f_\gamma c_{\alpha \gamma}-\sum_{\beta \gamma} f_ \beta f_\gamma c_{\beta \gamma}-c_{\alpha
\omega })$, and $\lamRs$ is the second-largest eigenvalue of the weighted square of the reader $R_{ij}=(u_{i}u_{j})^{1/2}
\sum_{k}(r_{ik}r_{kj}/\sum_{t}u_{t}r_{tk})$.

In the case of 2-by-2 code (\ref{IDsimple}),  $\lamRs = (1-\epsilon)^2$ and $\lamCs = c_0/2$, which by substitution in (\ref{kappa_c}) yield the critical value (\ref{kappac0}).  It is interesting to note that the eigenvalue $\lamRs$  corresponds to the \textbf{smoothest} non-uniform eigenvector $\delta\eai^\ast \neq 0$, which represents a coding state. This coding eigenvector is the \emph{first-excited }state of the system, which measures the preference of meanings to be encoded by specific symbols (see \cite{Tlusty07Math,Tlusty07JTB,Tlusty08PRL,Tlusty08PhysBio,Tlusty08PNAS} and references therein).

Our discussion so far assumed that the evolution of the code more or less follows the track of the optimal code, $e^\ast$. However, the coding system may get stuck at a metastable, sub-optimal state due to the ruggedness of the fitness landscape. There may exist, somewhere in the code fitness landscape, a superior, global optimum. Nevertheless, reaching this optimum is vary hard and requires crossing  deep `valleys' or following very intricate pathways. This system may exhibit slow, `glassy' dynamics. In this kind of almost frozen dynamics \cite{Sella06}, the local landscape is much more important than the location of the global optimum. In addition, other effects of population dynamics, such as mutations and genetic drift, may drive the coding system towards suboptimal states. Mutations broaden the population, creating a ``quasi-species" with a reduced effective fitness. Genetic drift delays the coding transition to higher gains (for further details see \cite{Tlusty08PRL,Tlusty08PhysBio,Tlusty08PNAS}).

\section{The topology of the symbol space and the coloring problem}\label{topology}
As mentioned above, the reader $r$ may be depicted in terms of a graph $G(r,\cS)$, which represents the topology of the symbol space $\cS$ by drawing an edge between every pair of symbols that are likely to be confused. The second-largest eigenvalue of the reader squared, $\lamRs$, which corresponds to the coding state, bears a special significance: The reader $r$ is related to the Laplacian of the symbol space $\Delta_{\cS}$ via  $\Delta_{\cS} = I - r$, where $I$ is the identity matrix \cite{Tlusty07JTB,Tlusty07Math,Tlusty08PRL}. Therefore, $\lamRs$ corresponds to the second-smallest eigenvalue $\lamDels$ of the Laplacian, $\lamRs = (1  - \lamDels)^2$ (in the degenerate 2-by-2 code, the second-smallest eigenvalue, $\lamDels =  \epsilon$, is its only available excited-state). The Laplacian operator appears naturally in the coding problem since it is the operator that describes random walk on the symbol graph  $G(r,\cS)$ via misreading events that move the molecular reader along edges connecting confused symbols. In fact, the eigenvalue $\lamDels$ is  the slowest relaxation time-scale of the system. The corresponding eigenvector $\delta\eai^\ast$ is known to be the smoothest of all excited modes of the graph, in accord with the intuitive physical notion that the modes with the lowest ``energy" eigenvalues and frequencies are those of the largest wave-lengths.

It follows from Courant's theorem that the smooth, first excited mode $\delta\eai^\ast$ divides the graph into two contiguous positive and negative regions  \cite{Tlusty07JTB,Tlusty07Math}. In the positive region, $\delta\eai^\ast \ge 0 $, the symbols will tend to encode certain meanings, whereas in the other region the chance to encode these meanings will be lower than the average, $\delta\eai^\ast \le 0 $. Thus, $\delta\eai^\ast$ partitions the graph with minimal boundaries between regions of opposite tendency to encode certain meanings and the emergent code is smooth, in the sense that \emph{adjacent symbols tend to encode similar meanings}. This arrangement minimizes the distortion $D$ by decreasing the average distance $c$ between meanings encoded by adjacent symbols. For example, if a coding system has two possible meanings, say `sea' and `land', then it is clear that the distortion of an arrangement according to the lowest-excited mode, where there is one continent and one ocean, is much smaller than the distortion of an intricate arrangement with many islands, seas, peninsulas and bays. Indeed, in the case of the genetic code, all amino acids are encoded by synonymous codons arranged in contiguous domains except serine that splits into two domains \cite{Crick68,Tlusty07JTB}. Our model concludes that the genetic code is smooth because the lowest-excited modes at the transition are the smoothest non-uniform modes. Similar continuity is found in the transcription regulation network \cite{Itzkovitz06,Shinar06}.

The low modes partition the symbol graph $G(r,\cS)$ into domains, which may be likened to drawing borders between countries on a map. We have found that the problem of maximizing the fitness of the code by optimizing this partition is related to another classical partition problem, the coloring problem \cite{Tlusty07JTB,Tlusty07Math,Tlusty08PNAS}. In the coloring problem, the goal is to calculate the minimal number of colors required to color an arbitrary map on a surface such that no two bordering countries have the same color. This minimal number is termed the ``coloring number" of the surface and is determined by the surface topology. It follows from our model that the topology of the code sets the coloring number as an upper limit to the number of first excited modes, and thus to the number of encoded  meanings. The relation of the coloring problem to the maximal number of first excited modes has a geometrical origin which is discussed in detail in \cite{Tlusty07JTB,Tlusty07Math}. For the genetic code, the coloring number estimate is in range of  $20-25$, in the neighborhood of the naturally occurring number. In the transcription regulation network \cite{Itzkovitz06}, the coloring number sets bounds that are close to the size of certain transcription factor families.

\section*{Acknowledgment}
The work was partially supported by the Minerva foundation, the Israel Science Foundation and the Clore Center. 

\bibliographystyle{IEEEtran}
\bibliography{phase}

\begin{thebibliography}{10}
\providecommand{\url}[1]{#1}
\csname url@samestyle\endcsname
\providecommand{\newblock}{\relax}
\providecommand{\bibinfo}[2]{#2}
\providecommand{\BIBentrySTDinterwordspacing}{\spaceskip=0pt\relax}
\providecommand{\BIBentryALTinterwordstretchfactor}{4}
\providecommand{\BIBentryALTinterwordspacing}{\spaceskip=\fontdimen2\font plus
\BIBentryALTinterwordstretchfactor\fontdimen3\font minus
  \fontdimen4\font\relax}
\providecommand{\BIBforeignlanguage}[2]{{%
\expandafter\ifx\csname l@#1\endcsname\relax
\typeout{** WARNING: IEEEtran.bst: No hyphenation pattern has been}%
\typeout{** loaded for the language `#1'. Using the pattern for}%
\typeout{** the default language instead.}%
\else
\language=\csname l@#1\endcsname
\fi
#2}}
\providecommand{\BIBdecl}{\relax}
\BIBdecl

\bibitem{Crick68}
F.~Crick, ``The origin of the genetic code.'' \emph{J. Mol. Biol.}, vol.~38,
  no.~3, pp. 367--79, 1968.

\bibitem{Tlusty07JTB}
T.~Tlusty, ``A model for the emergence of the genetic code as a transition in a
  noisy information channel,'' \emph{J. Theor. Biol.}, vol. 249, no.~2, pp.
  331--342., 2007.

\bibitem{Tlusty07Math}
------, ``A relation between the multiplicity of the second eigenvalue of a
  graph {L}aplacian, {C}ourant's nodal line theorem and the substantial
  dimension of tight polyhedral surfaces.'' \emph{Elec J Linear Algebra},
  vol.~16, pp. 315--324, 2007.

\bibitem{Tlusty08PRL}
------, ``Rate-distortion scenario for the emergence and evolution of noisy
  molecular codes,'' \emph{Phys. Rev. Lett.}, vol. 100, no.~4, pp. 048\,101--4,
  2008.

\bibitem{Tlusty08PhysBio}
------, ``A simple model for the evolution of molecular codes driven by the
  interplay of accuracy, diversity and cost,'' \emph{Physical Biology}, vol.~5,
  no.~1, p. 016001, 2008.

\bibitem{Tlusty08PNAS}
------, ``Casting polymer nets to optimize noisy molecular codes,'' \emph{Proc.
  Natl. Acad. Sci. U. S. A.}, vol. 105, no.~24, pp. 8238--8243, 2008.

\bibitem{Savir07}
Y.~Savir and T.~Tlusty, ``Conformational proofreading: the impact of
  conformational changes on the specificity of molecular recognition,''
  \emph{PLoS ONE}, vol.~2, p. e468, 2007.

\bibitem{Savir08}
------, ``Optimal design of a molecular recognizer: Molecular recognition as a
  {B}ayesian signal detection problem,'' \emph{IEEE Journal of Selected Topics
  in Signal Processing}, vol.~2, no.~3, pp. 390--399, 2008.

\bibitem{Savir09}
------, ``{DNA} extension induced by {R}ec{A} during homologous recombination
  as a possible conformational proofreading mechanism,'' \emph{submitted}.

\bibitem{Shinar06}
G.~Shinar, E.~Dekel, T.~Tlusty, and U.~Alon, ``Rules for biological regulation
  based on error minimization,'' \emph{Proc. Natl. Acad. Sci. U. S. A.}, vol.
  103, no.~11, pp. 3999--4004, 2006.

\bibitem{Itzkovitz06}
S.~Itzkovitz, T.~Tlusty, and U.~Alon, ``Coding limits on the number of
  transcription factors,'' \emph{BMC Genomics}, vol.~7, no.~1, p. 239, 2006.

\bibitem{Berger71}
T.~Berger, \emph{Rate distortion theory}.\hskip 1em plus 0.5em minus
  0.4em\relax NJ: Prentice-Hall, 1971.

\bibitem{Luttrell94}
S.~P. Luttrell, ``A {B}ayesian-analysis of self-organizing maps,'' \emph{Neural
  Comput.}, vol.~6, no.~5, pp. 767--794, 1994.

\bibitem{Graepel97}
T.~Graepel, M.~Burger, and K.~Obermayer, ``Phase transitions in stochastic
  self-organizing maps,'' \emph{Phys. Rev. E.}, vol.~56, no.~4, pp. 3876--3890,
  1997.

\bibitem{Rose98}
K.~Rose, ``Deterministic annealing for clustering, compression, classification,
  regression, and related optimization problems,'' \emph{Proc IEEE}, vol.~86,
  no.~11, pp. 2210--2239, 1998.

\bibitem{Tishby99}
N.~Tishby, F.~Pereira, and W.~Bialek, ``The information bottleneck method,'' in
  \emph{The 37-th Allerton Conf. on Communication, Control and Computing},
  1999, pp. 368--377.

\bibitem{Shannon59}
C.~E. Shannon, ``Coding theorems for a discrete source with a fidelity
  criterion,'' in \emph{National Convention Record}, vol. 7 (part 4), 1959, pp.
  142--163.

\bibitem{Cover06}
T.~M. Cover and J.~A. Thomas, \emph{Elements of information theory},
  2nd~ed.\hskip 1em plus 0.5em minus 0.4em\relax Hoboken, NJ:
  Wiley-Interscience, 2006.

\bibitem{Sella06}
G.~Sella and D.~Ardell, ``The coevolution of genes and genetic codes: Crick’s
  frozen accident revisited,'' \emph{J. Mol. Evol.}, vol.~63, no.~3, pp.
  297--313, 2006.

\end{thebibliography}

\end{document}